\documentclass{ws-ijgmmp}

\usepackage{graphicx}  
\usepackage{latexsym}
\usepackage{amssymb}

\def\beq{\begin{equation}}
\def\eeq{\end{equation}}

\begin{document}

\markboth{D. Bini, F. de Felice and A. Geralico}
{Observer-dependent optical properties of stationary axisymmetric spacetimes}

%
\catchline{}{}{}{}{}
%

\title{OBSERVER-DEPENDENT OPTICAL PROPERTIES OF STATIONARY AXISYMMETRIC SPACETIMES
}

\author{\footnotesize DONATO BINI}

\address{
Istituto per le Applicazioni del Calcolo ``M. Picone,'' CNR, I--00185 Rome, Italy\\
ICRA, ``Sapienza'' University of Rome,  I--00185 Rome, Italy\\
INFN, Sezione di Firenze, I--00185 Sesto Fiorentino (FI), Italy
\email{binid@icra.it}
}

\author{\footnotesize FERNANDO DE FELICE}

\address{
Department of Physics and Astronomy \lq\lq G. Galilei," University of Padova, via Marzolo 8, I-35131 Padova, Italy
\email{defelicef43@gmail.com}
}

\author{ANDREA GERALICO}

\address{
Physics Department and ICRA, ``Sapienza'' University of Rome, I--00185 Rome, Italy
\email{geralico@icra.it}
}

\maketitle

\begin{history}
\received{(Day Month Year)}
\revised{(Day Month Year)}
\end{history}

\begin{abstract}
The world lines of null particles admit arbitrary parametrizations. In the presence of a family of observers one may introduce along a null world line an extension of the so-called Cattaneo's relative standard time parameter (valid for massive particles) which plays a special role. Another possibility is to use the coordinate time itself as a parameter.
The relation between relative standard time and coordinate time allows for the introduction of an observer-dependent  optical path and associated refraction index. Both these quantities are studied here working out explicit examples concerning familiar null orbits and observers in black hole spacetimes.
\end{abstract}


\section{Introduction}

As it is well known, null world lines admit arbitrary parametrizations.
However, in the presence of a family of observers defined all along a null world line,
one may introduce as a parameter on that world line the value of Cattaneo's relative standard time \cite{cattaneo}, which would be read on the clock of the local observer being crossed by the null line. 
Similarly, another special parameter may be related to the used coordinate patch and in particular to the coordinate time itself.

Generalizing a pioneering approach due to M\"oller \cite{moller}, the relation between relative standard time and coordinate time is used here to introduce an observer-dependent optical path with an associated refraction index. Their features are studied in detail also exploring the transformation properties when we change observers as well as coordinates. 

It is worth to mention that this formulation is closely related to the action integral leading to Fermat's principle in general relativity (see, e.g., Ref. \cite{perlick}).
In ordinary optics, it provides a powerful tool to deal with different kinds of material media with varying refraction index and to study the properties of light rays.
Moreover, an observer-dependent refraction index has been already defined in Ref. \cite{mfg}, but only in the case of special observes, namely the static ones, having their four velocity aligned with the coordinate temporal lines, and the locally nonrotating ones, having their four velocity orthogonal to constant coordinate time hypersurfaces. Considering the case of general observers is then an original contribution of this work: our approach can be exploited to look for other observers who play a role in this context.

We discuss here explicit examples concerning familiar null orbits and observers in stationary axisymmetric spacetimes.
For instance, we show that in a Kerr background the so-called Painlev\'e-Gullstrand observers \cite{Painleve21,Gullstrand22}, i.e., the family of geodesic observers moving radially with respect to the locally nonrotating ones, have unit relative refraction index irrespective of the photon path if the metric is written in adapted (horizon-penetrating) coordinates, a fact that further specifies their already known geometrical and physical properties.

Units are such that $G=1$ and $c=1$, $G$ being the Newtonian constant and $c$ the speed of light in vacuum. Greek indices run from $0$ to $3$ and latin indices run from $1$ to $3$. The metric signature is chosen to be $+2$.

\section{Null orbits and physical observers in a general spacetime: observer-dependent optical path and refraction index} 

Consider a coordinate system $x^\alpha=\{ t, x^a\}$ and a general spacetime metric $d s^2=g_{\alpha\beta}d x^\alpha d x^\beta$ with signature $-+++$, as stated.
Let $u$ be the four velocity field of a family of observers defined in a certain spacetime region (i.e., $u$ is the future-pointing unit vector tangent to the timelike world lines of the observers) and $P(u)=g+u\otimes u$ the projection operator in the Local Rest Space of $u$ ($LRS_u$).
The observer-adapted (local) representation of the spacetime metric is then
\beq
g=-u\otimes u + P(u)\,,
\eeq
with associated line element
\beq
\label{met_u}
ds^2=-(u_\alpha dx^\alpha)^2+P(u)_{\alpha\beta}dx^\alpha dx^\beta\,. 
\eeq
Let $P^\alpha=dx^\alpha/d\lambda$ be the vector field tangent to a null orbit ${\mathcal C}_{(P)}$, not necessarily a geodesic,\footnote{See, e.g., Exercise 33.3 of Ref. \cite{mtw} for photons constrained along a circular optical guide.
} $\lambda$ being an arbitrary affine parameter along it.
With respect to the chosen  observers $u$, the photon four momentum writes as
\beq
\label{Pdef}
P=P^\alpha \partial_\alpha=E(P,u)[u+\hat \nu(P,u)]\,,
\eeq
where $\hat \nu(P,u)$ is a unit spatial vector in the $LRS_u$, with $\hat \nu(P,u)\cdot \hat \nu(P,u)=1$,  representing the
local line of sight (the spatial photon direction) of the given observers $u$ and $E(P,u)$ is the photon's relative energy.
In spite of the arbitrariness of $\lambda$, the presence of the family of observers induces  a special parametrization
along ${\mathcal C}_{(P)}$ which is an extension of the so-called Cattaneo relative standard time \cite{mfg}, namely
\beq
\label{prop}
d\tau_{(P,u)}=-u^\flat |_{{\mathcal C}_{(P)}}
=-u_\alpha dx^\alpha|_{{\mathcal C}_{(P)}}=-u_\alpha \frac{dx^\alpha}{d\lambda}\, d\lambda=E(P,u)\, d\lambda\,,
\eeq
where the symbol $\flat$ indicates the fully covariant form of a tensor.
Similarly, the quantity
\beq
\label{dell_def}
d\ell_{(P,u)}=\sqrt{P(u)_{\alpha\beta}dx^\alpha dx^\beta}\, \bigg|_{{\mathcal C}_{(P)}}
\eeq
is an observer-dependent length element (relative standard length) which 
represents a natural choice of the  parameter on the null orbit.
From the vanishing of the spacetime interval given by Eq. (\ref{met_u}) along a null orbit, one has
\beq
0=-d\tau_{(P,u)}^2+ d\ell_{(P,u)}^2\,,
\eeq
that is
\beq
\label{d_tau}
d\tau_{(P,u)} =d\ell_{(P,u)} \,,
\eeq
where the choice of the $+$ sign in Eq. (\ref{d_tau}) implies that  $d\ell_{(P,u)}$ increases with $d\tau_{(P,u)}$. 
As stated, the property of $\ell_{(P,u)}$ being  an  arclength is assured by the relation
\beq
\label{lprop}
\sqrt{P(u)_{\alpha\beta}\frac{dx^\alpha}{d\ell_{(P,u)}}\frac{dx^\beta}{d\ell_{(P,u)}}}=1\,.
\eeq
From Eq. (\ref{d_tau}) it follows that the rate of increase of the parameter $\ell_{(P,u)}$ along the null orbit coincides with that of $\tau_{(P,u)}$ along the observer world line. This is the relation one would expect for a light ray in a Minkowski spacetime. 

One may also parametrize the photon world line by using the coordinate time, all the parameters being simply related
\beq
\label{prop2}
d\tau_{(P,u)}=d\ell_{(P,u)}=E(P,u)\, d\lambda=\frac{E(P,u)}{P^t}\, dt \,.
\eeq
The latter coordinate time parametrization of the photon world line may be used to define an \lq\lq observer-dependent optical path"
which reminds the classical expression and leads to a closely related definition of \lq\lq observer-dependent refraction index," in the sense specified below.

Starting from Eqs. (\ref{prop}) and (\ref{d_tau}) one finds
\beq
\label{dell_def2}
-u_\alpha dx^\alpha=d\ell_{(P,u)}\quad \to \quad
-u_t dt=d\ell_{(P,u)}+u_a dx^a\,,
\eeq
so that, provided $u_t\not =0$
\beq
\label{eq:2.10}
dt=-\frac{d\ell_{(P,u)}}{u_t} \left(1+u_a T_u^a \right)\,, \qquad
T_u^a=\frac{dx^a}{d\ell_{(P,u)}}\,.
\eeq
Therefore, once a specific choice is made of the coordinates, the family of observers and the photon orbit (together with its momentum), then Eq. (\ref{eq:2.10}) gives the relation between the observer relative length parametrization along the null world line and the coordinate time.
The relations in (\ref{eq:2.10}) can be cast in the optical-like form (see, e.g., Ref. \cite{mfg} and references therein)
\beq
\label{coord_time}
dt=n_{u} d\ell_{(P,u)} \,,\qquad n_u= -\frac{1}{u_t} \left(1+u_a T_u^a\right)\,,
\eeq
leading to a (local) observer-dependent refraction index, $n_u$, only defined, however, along the null world line.
It is interesting to study under what conditions $n_u\le1$, i.e.,
\beq
-\frac{1}{u_t} \left(1+u_aT_u^a\right)\le 1\,.
\eeq
In order to have $u$ future-pointing ($u^t>0$), $u_t$ must be negative and hence the previous relation becomes
\beq
\label{n_less_1_cond}
1+u_aT_u^a \le |u_t|\quad \to \quad u_a T_u^a\le |u_t|-1\,.
\eeq
In general, the refraction index is bigger than 1 and this means, from Eq. (\ref{n_de_l}),  that $dt>  d\ell_u$, being $d\ell_u>0$; namely, 
the coordinate distance $dt$ covered by the light ray is longer than the optical (physical) distance, i.e.,  that 
covered by a light ray moving through a medium (the medium is optically dense).
A refraction index $n_u< 1$, instead, means that the relative arclength increases faster than  the coordinate time ($dt<d\ell_u$); namely, 
the light signal would move in the medium faster than it would do in the absence of it, a fact that may give rise to counter-intuitive effects. When $n_u=1$, 
the light ray behaves as if it were in the absence of the physical medium and in this case we shall say that the time coordinate is adapted to the observers.

Note that the refraction index (\ref{eq:2.10}) is not covariantly defined, arising from metric components and transforming correspondingly;
therefore, a change of coordinates (time in particular) would change it, even
keeping the observer as fixed. This is not surprising if one interprets a change of coordinates as a change
of \lq\lq perspective.''
In Section \ref{PGobs} we will provide an example of the effect of a change of coordinates on the refraction index associated with a selected family of observers.

On the other hand, within a given set of coordinates and for a fixed photon path one may change the observer, passing for instance from $u$ to any other observer $U$, in relative motion with respect to $u$.
The new observer $U$ will then be associated with a different refraction index $n_U$, namely
\beq
\label{n_de_l}
dt=n_{u} d\ell_{(P,u)}=n_{U} d\ell_{(P,U)}\,,
\eeq
implying in turn the transformation law 
$n_u=n_U {d \ell_{(P,U)}}/{d \ell_{(P,u)}}$\,.

\subsection{Special observers}

The above results for the local definition of refraction index (\ref{coord_time}) can be conveniently specialized to a particular family of observers.
In any spacetime there exist families of observers which are naturally related to a given choice of coordinates and therefore  play a central role. Here we refer to those observers having world lines aligned with the time coordinate lines and those having world lines orthogonal to constant time hypersurfaces:

\begin{enumerate}

\item {\it Static observers}\\

These are at rest with respect to the space coordinates with four velocity tangent to the coordinate time axis
\beq
m=\frac{1}{M}\partial_t \,,\qquad m^\flat=-M(dt-M_a dx^a)\,, 
\eeq
where $M=\sqrt{-g_{tt}}$ (lapse function), $M_a=-g_{ta}/g_{tt}$ (shift function). Denoting the spatial metric as $\gamma_{ab}=g_{ab}+M^2M_aM_b$,
the form of the spacetime metric adapted to these observers is  given by
\beq
\label{metthd}
ds^2=-M^2(dt-M_a dx^a)^2+\gamma_{ab}dx^adx^b\,.
\eeq 
Relative to such observers the refraction index (\ref{coord_time}) is 
\beq
\label{nstatic}
n_m=\frac{1}{M} +M_a T_m^a\,, \qquad T_m^a=\frac{dx^a}{d\ell_m}\,,
\eeq
for a general null trajectory.

\item {\it Locally nonrotating observers (lnor)}\\

These have four velocity orthogonal to the (coordinate) $t=const.$ hypersurfaces
\beq
\label{zamos_varie}
n=\frac{1}{N}\left(\partial_t-N^a \partial_a\right) \,,\qquad n^\flat=-Ndt\,, 
\eeq
where $N=1/\sqrt{-g^{tt}}$ and $N_a=g_{ta}$. 
The form of the metric adapted to these observers is  given by
\beq
\label{metsli}
ds^2=-N^2dt^2+g_{ab}(dx^a+N^a dt)(dx^b+N^b dt)\,.
\eeq 
For them, the refraction index (\ref{coord_time}) is 
\beq
\label{nrefzamo}
n_n=\frac{1}{N}\,;
\eeq
as we can see, $n_n$ is independent of the photon momentum, a property which further characterizes the lnor ($n$) observers, in addition to their defining properties of being tangent to a vorticity-free congruence and having zero angular momentum component along $\phi$, $n_\phi=0$.
It is interesting to consider lnor observers and adapted coordinates so that the lapse function $N=1$, implying, as a consequence, that the refraction index is always $n_n=1$.
This situation occurs for example in spacetimes admitting separable geodesics \cite{geosli}.
In particular, this is the case of Painlev\'e-Gullstrand observers \cite{Painleve21,Gullstrand22} in the Kerr family of spacetimes and coordinates adapted to them.

\end{enumerate}

We will now proceed  by considering some explicit spacetimes referring to both static and locally nonrotating observers as well as to other relevant family of observers.

\section{Stationary and axisymmetric spacetimes}

In a stationary and axisymmetric spacetime with coordinates $x^0=t$ and $x^3=\phi$ adapted to the stationary and rotational Killing directions, the line element can be written as
\begin{eqnarray}
\label{axisym}
ds^2&=&-M^2(dt-M_\phi d\phi)^2+\gamma_{ab}dx^adx^b\nonumber\\
&=&-N^2dt^2+g_{ab}(dx^a+N^\phi dt)(dx^b+N^\phi dt)\,,
\end{eqnarray}
with the metric functions not depending on $t$ and $\phi$ (the not ignorable coordinates $x^1$ and $x^2$ will be specified later).
Here we have
\beq
M^2=N^2-N_\phi N^\phi\,,\ \quad M^2M_\phi=N_\phi\,,\quad \gamma_{ab}=g_{ab}+M^2M_\phi^2 \delta_a^\phi \delta_b^\phi\,.
\eeq

Due to the rotational symmetry of the spacetime metric, observers moving along spatially circular orbits play a special role.
Their four velocity is given by
\beq
\label{ucirc}
u=\Gamma(\partial_t + \zeta\partial_\phi)\,,  \qquad 
\Gamma=[-(g_{tt}+2\zeta g_{t\phi}+\zeta^2 g_{\phi\phi})]^{-1/2}\,,
\eeq
with fully covariant representation
\beq
u^\flat=u_t dt +u_\phi d\phi=\Gamma[(g_{tt}+\zeta g_{t\phi})dt+(g_{t\phi}+\zeta g_{\phi\phi})d\phi]\,.
\eeq
Notice that when $\zeta=0$ we recover the case of a static observer, whereas when $\zeta=-N^\phi$ we recover the case of locally nonrotating observers (if the shift has the only nonvanishing component $N^a=N^\phi\delta^a_\phi$).
For circularly rotating observers the local refraction index reads
\beq
\label{n_u_circ}
n_{u}= -\frac{1}{u_t} \left(1+u_\phi T_u^\phi  \right)= -\frac{1}{u_t} \left(1+u_\phi \frac{d\phi}{dt}\frac{dt}{d\ell_u}  \right)\,, \qquad
T_u^\phi=\frac{d\phi}{d\ell_u}\,,
\eeq
that is
\beq
n_{u}=-\frac{1}{u_t} \left(1+u_\phi \zeta_{\rm (c)} n_{u}\right)\,,
\eeq
having introduced the photon angular velocity $\zeta_{\rm (c)} ={d\phi}/{dt}$.
Isolating $n_{u}$ we obtain
\beq
\label{nucirc}
n_{u}=-\frac{1}{\left(u_t+ u_\phi \zeta_{\rm (c)} \right)} \,.
\eeq

As an application, consider in the equatorial plane of a central source a pair of  photons moving on spatially circular orbits and in opposite directions to each other.
The associated four momentum of the circular (co-rotating, $+$, and counter-rotating, $-$) null orbit is
\beq
P_\pm=\Gamma_{\pm} (\partial_t + \zeta_{\pm} \partial_\phi)\,,\qquad 
\frac{1}{\zeta_\pm}=M_\phi \pm \frac{\sqrt{\gamma_{\phi\phi}}}{M}\,,
\eeq
where $\Gamma_{\pm}=dt_\pm /d\lambda$ is an arbitrary parameter (allowed by the freedom in the choice of the photon orbit parametrization) and $x^1=$ const., $x^2=$ const..

In the case of static observers, the refraction index along co-rotating and counter-rotating null orbits is given by Eq. (\ref{nstatic}) with $T^a_m=\delta^a_\phi d\phi/d\ell_m$.
Using Eq. (\ref{lprop}), i.e.,  
\beq
d\ell_{(P_\pm ,m)}=\sqrt{\gamma_{\phi\phi}}|d\phi|=\pm\sqrt{\gamma_{\phi\phi}}d\phi\,,
\eeq
the refraction index reduces to 
\beq
\label{nrefstat}
n_m^\pm=\frac{1}{M}+M_{\phi}\frac{d\phi}{d\ell_{(P_\pm ,m)}}
=\frac{1}{M}\pm \frac{M_{\phi}}{\sqrt{\gamma_{\phi\phi}}}
=\pm\frac{1}{\sqrt{\gamma_{\phi\phi}}}\frac{1}{\zeta_\pm}
=\frac{1}{\sqrt{\gamma_{\phi\phi}}}\frac{1}{|\zeta_\pm|}\,.
\eeq
A nice geometrical property of $n_m^\pm$ is that it can be expressed in terms of the period ${\mathcal T}_m^{\pm}={2\pi}/{|\zeta_\pm|}$ and the proper circumference ${\mathcal L}_m=2\pi \sqrt{\gamma_{\phi\phi}}$ of the associated orbit, as follows
\beq
\label{bella}
n_m^\pm=\frac{1}{2\pi \sqrt{\gamma_{\phi\phi}}}\, \frac{2\pi}{|\zeta_\pm|}=\frac{{\mathcal T}_m^{\pm}}{{\mathcal L}_m}\,.
\eeq
Requiring $n_m^\pm<1$ ($n_m^\pm>1$) implies that ${\mathcal T}_m^{\pm}<{\mathcal L}_m$ (${\mathcal T}_m^{\pm}>{\mathcal L}_m$), whereas the special case $n_m^\pm=1$  (whenever possible) corresponds to what we expect in the optical medium approach consistently with the Euclidean assertion that
the length of the orbit equals its period.  Note that in Fig. \ref{fig_n_thd_sli_kerr} for the specific case of a Kerr spacetime we see that only $n_m^+$ can be set as 1.
Furthermore, if ${\mathcal L}_m/{\mathcal T}_m^{\pm}={\mathcal V}_m^\pm$ denotes a \lq\lq mean relative spatial velocity," then  Eq. (\ref{bella}) gives exactly ${\mathcal V}_m^\pm=1/n_m^\pm$, as expected from the optical analogy.   

In the case of locally nonrotating observers, we have
\beq
n=\frac{1}{N}(\partial_t + \zeta_{\rm (lnor)}\partial_\phi)\,,\qquad \zeta_{\rm (lnor)}=-N^\phi\,,
\eeq
and the refraction index is simply given by Eq. (\ref{nrefzamo}), irrespective of the chosen photon path.
Also in this case one has the nice geometrical interpretation in terms of the period ${\mathcal T}_n={2\pi}/{|\zeta_\pm-\zeta_{\rm (lnor)}|}$ and the proper circumference ${\mathcal L}_n=2\pi \sqrt{g_{\phi\phi}}$ of the associated orbit, namely
\beq
n_n=\frac{1}{N}=\frac{1}{2\pi \sqrt{g_{\phi\phi}}}\, \frac{2\pi}{|\zeta_\pm-\zeta_{\rm (lnor)}|}=\frac{{\mathcal T}_n}{{\mathcal L}_n}=\frac{1}{{\mathcal V}_n}\,,
\eeq
which is the same for both co-rotating and counter-rotating photons \cite{idcf2}, being
\beq
\zeta_+-\zeta_{\rm (lnor)}=-(\zeta_--\zeta_{\rm (lnor)})\,.
\eeq
This result is expected because the \lq\lq local nonrotation" of the chosen observers has the consequence of canceling the effect 
of the spacetime rotation and dragging.

For observers moving along spatially circular orbits with four velocity $u$ (see Eq. (\ref{ucirc})) it is convenient to express it in terms of the linear velocity $\nu$ as follows
\beq
u=\gamma(n+\nu e_{\hat\phi})\,, \qquad 
\gamma=\frac{1}{\sqrt{1-\nu^2}}\,,
\eeq
with $n$ representing the lnor family of fiducial observers, $\nu\in(-1,1)$ related to the angular velocity $\zeta$ by
\beq
\nu=(\zeta-\zeta_{\rm(lnor)})\frac{\sqrt{g_{\phi\phi}}}{N}\,,
\eeq
and $\gamma=N\Gamma$ the Lorentz gamma factor of $u$ with respect to $n$.
We recall that $N$ is the lapse function defined after Eq. (\ref{zamos_varie}) 
and $e_{\hat \phi}=(1/{\sqrt{g_{\phi\phi}}})\partial_\phi$ denotes the unit vector along the azimuthal direction. 

One can then generalize the above results by evaluating the period of revolution of the null orbit as well as the
 proper length of its trajectory according to the relations
\begin{eqnarray}
{\mathcal T}_u^{\pm}&=&\frac{2\pi}{|\zeta_{\pm}-\zeta|}=2\pi\frac{\sqrt{g_{\phi\phi}}}{N}\frac{1}{1\pm\nu}={\mathcal T}_n \frac{1}{1\pm \nu}\,, \nonumber\\
{\mathcal L}_u&=&2\pi\sqrt{P(u)_{\phi\phi}}=2\pi\sqrt{g_{\phi\phi}+u_{\phi}^2}=2\pi\sqrt{g_{\phi\phi}}\gamma={\mathcal L}_n\gamma\,.
\end{eqnarray}
It turns out that the refraction index (\ref{nucirc}) can be written in terms of $\nu$ as
\beq
\label{nucirc2}
n_{u}^{\pm}=\frac{{\mathcal T}_u^{\pm}}{{\mathcal L}_u}\equiv\frac{1}{{\mathcal V}_u^\pm}
=\frac{1}{{\mathcal V}_n}\sqrt{\frac{1\mp\nu}{1\pm\nu}}
=n_n\sqrt{\frac{1\mp\nu}{1\pm\nu}}\,,
\eeq
or, explicitly, for co-rotating and counter-rotating orbits
\beq
n_{u}^{+}=n_n\sqrt{\frac{1-\nu}{1+\nu}}\,, \qquad
n_{u}^{-}=n_n\sqrt{\frac{1+\nu}{1-\nu}}\,,
\eeq
with the nice property $n_{u}^{+}n_{u}^{-}=n_u^2$.

\begin{figure} 
\begin{center}
\includegraphics[scale=0.4]{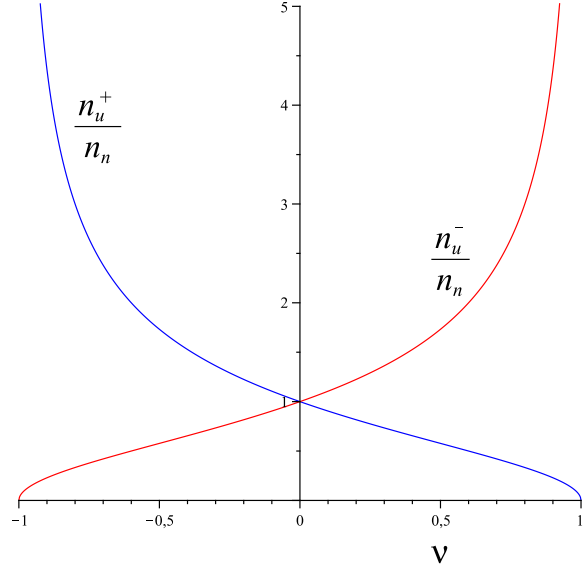}
\end{center}
\caption{ 
The behaviors of the ratios between the refraction indices $n_{u}^{\pm}$ associated with circularly rotating observers and $n_n$ are shown as functions of $\nu$. 
}
\label{fig:nupm_su_nun_vs_nu}
\end{figure}

As expected, a refraction index less than 1 implies an effective velocity ${\mathcal V}_u^\pm$ larger than 1, a situation which should be disregarded as unphysical, unless one concludes that in the above conditions the gravitational field behaves as an exotic medium.
Nevertheless, it is always possible to identify a family of circularly rotating observers with respect to whom the refraction index associated with co/counter-rotating photons appears equal to unity. In this case, solving Eq. (\ref{nucirc2}) for $\nu$, we obtain
\beq
\label{nustar}
\nu_*=\pm\frac{1-{\mathcal V}_n^2}{1+{\mathcal V}_n^2}
=\pm\frac{1-N^2}{1+N^2}\,,
\eeq
where the sign choice is done accordingly to the specified co/counter-rotating null orbit.
The behaviors of the ratios between the refraction indices $n_{u}^{\pm}$ and $n_n$ as functions of $\nu$ are shown in Fig. \ref{fig:nupm_su_nun_vs_nu}.


\subsection{The Kerr spacetime}

Let us consider the case of the Kerr spacetime with metric written in standard Boyer-Lindquist coordinates $(t,r,\theta,\phi)$ in the form (\ref{axisym}) with 
\begin{eqnarray}
\label{kerrfunct1}
[M^2,M_\phi]&=&\left[1-\frac{2{\mathcal M}r}{\Sigma},-\frac{2a{\mathcal M}r\sin^2\theta}{\Sigma-2{\mathcal M}r}\right]\,,\nonumber\\{}
[N^2,N_\phi]&=&\left[\frac{\Delta\Sigma}{\Delta\Sigma+2{\mathcal M}r(r^2+a^2)},-\frac{2a{\mathcal M}r\sin^2\theta}{\Sigma}\right]\,,
\end{eqnarray}
and
\begin{eqnarray}
\label{kerrfunct2}
[\gamma_{rr},\gamma_{\theta\theta},\gamma_{\phi\phi}]&=&\Sigma\,\left[\frac{1}{\Delta},1,\frac{\Delta\sin^2\theta}{\Sigma-2{\mathcal M}r}\right]\,,\nonumber\\{}
[g_{rr},g_{\theta\theta},g_{\phi\phi}]&=&\Sigma\,\left[\frac{1}{\Delta},1,\frac{\sin^2\theta}{\Sigma^2}[\Delta\Sigma+2{\mathcal M}r(r^2+a^2)]\right]\,,
\end{eqnarray}
where $\Delta=r^2-2{\mathcal M}r+a^2$ and $\Sigma=r^2+a^2\cos^2\theta$.
Here $a$ and $M$ are the specific angular momentum and total mass of the metric source characterizing the spacetime. The event horizons are located at $r_\pm=M\pm\sqrt{M^2-a^2}$.
The angular velocities of the spatially circular null orbits in the equatorial plane $\theta=\pi/2$ are given by \cite{idcf2}
\beq
\label{zetapmnull}
\zeta_\pm=\frac{2a{\mathcal M}\pm r\sqrt{\Delta}}{r^3+a^2r+2a^2{\mathcal M}}\,.
\eeq

Relative to the static observers, the refraction index (\ref{nrefstat}) turns out to be
\beq
n_m^\pm=\frac{1}{\sqrt{r(r-2{\mathcal M})}}\left(r\mp\frac{2a{\mathcal M}}{\sqrt{\Delta}}\right)\,,
\eeq
where $r$ is the coordinate radius of the circular orbit.
Among all circular equatorial null orbits there exist only two null geodesics  which are located at the radii $r_{\rm (null,\pm)}$ solutions of the cubic equation
\beq
r(r-3{\mathcal M})\pm 2a{\mathcal M} \sqrt{\frac{r}{{\mathcal M}}}=0\,;
\eeq
the latter can be solved exactly to give for co-rotating $(+)$ and counter-rotating $(-)$ orbits
\beq
r_{\rm (null,\pm )}=2{\mathcal M} \left[1+\cos \left(\frac23 {\rm arccos}\left(\mp \frac{a}{{\mathcal M}} \right)  \right)  \right]\,.
\eeq
From these expressions, the refraction indices $n_m^\pm$ relative to the static observers can be written in terms of the solution parameters $(a,{\mathcal M})$ only, provided $a\le {\mathcal M}$.

\begin{figure} 
\begin{center}
\includegraphics[scale=0.4]{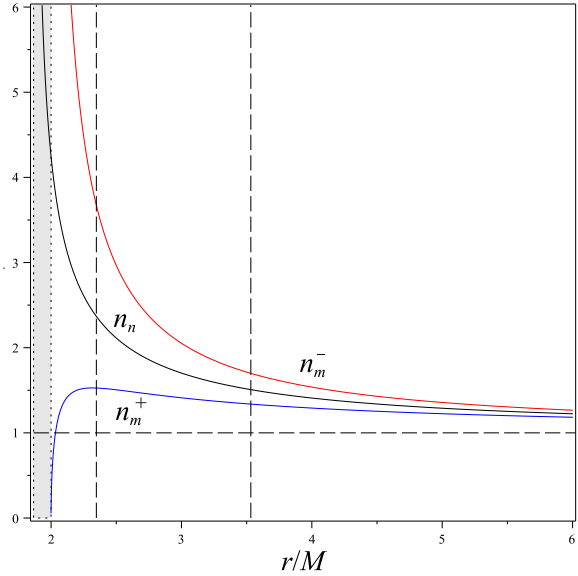}
\end{center}
\caption{
The behaviors of the refraction indices $n_m^\pm $ and $n_n$ for circular equatorial ($\theta=\pi/2$) null orbits in a Kerr spacetime are shown as functions of $r/{\mathcal M}$ for $a/{\mathcal M}=0.5$. Static observers exist for $r/{\mathcal M}>2$ only, i.e., outside the ergosphere (shaded region), whereas locally nonrotating ones are well defined all the way up to the horizon $r_+/{\mathcal M}\approx1.867$.
The values of the refraction indices for null geodesics correspond to the intersections of the corresponding curves with the dashed vertical lines at $r_{\rm (null,-)}/{\mathcal M}\approx2.347$ and $r_{\rm (null,+)}/{\mathcal M}\approx3.532$. At $r_*/{\mathcal M}\approx 2.034$ we find $n_m^+=1$, so that in the region $r_+/{\mathcal M}<r/{\mathcal M}<2.034$ we have that $n_m^+<1$.
}
\label{fig_n_thd_sli_kerr}
\end{figure}

The refraction index associated with locally nonrotating observers is instead given by Eq. (\ref{nrefzamo}), which now becomes
\beq
\label{niceprop}
n_n=\sqrt{1+\frac{2{\mathcal M}(r^2+a^2)}{r\Delta}}=\sqrt{n_m^+n_m^-}\,,
\eeq
allowing the identification of $n_n$ as the geometrical mean of $n_m^+$ and $n_m^-$. 
This nice property actually holds for any stationary axisymmetric spacetime, see below. 
The behaviors of the refraction indices $n_m^\pm $ and $n_n$ as functions of the radial coordinate are shown in Fig. \ref{fig_n_thd_sli_kerr}.

Static observers exist only outside the ergosphere, whose intersection with the equatorial plane is located at $r=2{\mathcal M}$. It turns out that the refraction index associated to co-rotating orbits goes to zero while approaching the ergosphere, whereas for counter-rotating orbits it diverges there.
In fact, for $r\to2{\mathcal M}$ the refraction indices $n_m^\pm $ behave as
\begin{eqnarray}
n_m^+&\sim&\frac{2{\mathcal M}^2+a^2}{\sqrt{2{\mathcal M}}a^2}\sqrt{r-2{\mathcal M}}+O[(r-2{\mathcal M})^{3/2}]\,,\nonumber\\
n_m^-&\sim&\frac{2\sqrt{2{\mathcal M}}}{\sqrt{r-2{\mathcal M}}}+O[(r-2{\mathcal M})^{1/2}]\,.
\end{eqnarray}
Furthermore, it is interesting to note that approaching the ergosphere the angular velocities (\ref{zetapmnull}) of the null orbits become
\beq
\lim_{r\to 2{\mathcal M}}\zeta_+=\frac{a}{(2{\mathcal M})^2+a^2}=\zeta_{\rm(car)}|_{r= 2{\mathcal M}} \,,\qquad \lim_{r\to 2{\mathcal M}}\zeta_-=0 \,,
\eeq
where $\zeta_{\rm(car)}=a/(r^2+a^2)$ is the angular velocity associated with Carter observers, whose virtue is that of  measuring  electric and magnetic fields parallel to each other \cite{carter}.
In contrast, locally nonrotating observers exist all the way down to the horizon, and the associated refraction index grows indefinitely there.
In the limit of a Schwarzschild spacetime ($a=0$) the refraction index turns out to be
\beq
n_m=\sqrt{\frac{r}{r-2{\mathcal M}}}=n_n\,,
\eeq
which diverges while approaching the horizon. An increasing refraction index means that the observer \lq\lq sees" the spacetime as a medium more and more dense. 
A diverging refraction index is then consistent with the complete lack of information reaching the observer.

Finally, the refraction indices associated with circularly rotating observers are given by Eq. (\ref{nucirc2}). 
Among them, the special family of observers seeing unit refraction index is characterized by
\beq
\nu_*=\pm\frac{{\mathcal M}(r^2+a^2)}{r(r^2+a^2)-{\mathcal M}(r^2-a^2)}\,,
\eeq
which is a monotonic function of the radial coordinate going to $\pm1$ approaching the horizon and vanishing at infinity.
In the Schwarzschild limit it becomes
\beq
\nu_*=\pm\frac{{\mathcal M}}{r-{\mathcal M}}\,.
\eeq


\subsection{Radial observers (with respect to lnor) and the Painlev\'e-Gullstrand family}
\label{PGobs}

We have considered so far special families of observers which are well suited to investigate both geometrical and physical properties of stationary axisymmetric spacetimes, i.e., the static observers, the locally nonrotating observers and the observers moving along spatially circular orbits.
There is a further family of observers, complementary in some sense to the circularly rotating ones, which plays a role in the Kerr  spacetime.
It is the family of observers moving radially with respect to lnor, i.e., with four velocity 
\beq
\label{PGlike}
{\mathcal N}=\gamma_{{\mathcal N}}[n+\nu({\mathcal N},n)^{\hat r}e_{\hat r}]\,, \qquad
{\mathcal N}^\flat= -dt +\frac{\sqrt{g_{rr}}}{N}\nu({\mathcal N},n)^{\hat r}dr\,,
\eeq
where 
$e_{\hat r}=(1/{\sqrt{g_{rr}}})\partial_r$ denotes the unit vector along the radial direction. 
For them, the local refraction index turns out to be given by
\beq
\label{n_u_rad}
n_{{\mathcal N}}= -\frac{1}{{\mathcal N}_t} \left(1+{\mathcal N}_r T_{\mathcal N}^r\right)
=-\frac{1}{{\mathcal N}_t} \left(1+{\mathcal N}_r \frac{dr}{dt}\frac{dt}{d\ell_{\mathcal N}}\right)
\,, \qquad
T_{\mathcal N}^r=\frac{dr}{d\ell_{\mathcal N}}\,,
\eeq
that is
\beq
\label{nurad}
n_{{\mathcal N}}=-\frac{1}{\left({\mathcal N}_t+ {\mathcal N}_r \zeta_{\rm (r)} \right)} \,,
\eeq
having introduced the photon radial velocity $\zeta_{\rm (r)} ={dr}/{dt}$. 
General null orbits on the symmetry plane, instead, correspond to
\beq
P_{\rm (gen)}=\Gamma_{\rm (gen)} \left[\partial_t +\zeta_{\rm (r)}\partial_r +\zeta_{\rm (c)}\partial_\phi  \right]
=E(P_{\rm (gen)},n)[n+\hat \nu^{\hat r}e_{\hat r}+\hat \nu^{\hat \phi}e_{\hat \phi}]\,,
\eeq
with $\hat \nu^{\hat r}{}^2+\hat \nu^{\hat \phi}{}^2=1$, or, equivalently,
\beq
\epsilon_r \sqrt{g_{rr}}\zeta_{\rm (r)}=[-(g_{tt}+2g_{t\phi}\zeta_{\rm (c)}+g_{\phi\phi}\zeta_{\rm (c)}^2)]^{1/2}\,.
\eeq

In stationary and axisymmetric spacetimes these observers form a hypersurface-orthogonal  congruence, i.e., their vorticity field identically vanishes. The additional request of being geodesic implies $\gamma_{{\mathcal N}}=1/N$ and $\nu({\mathcal N},n)^{\hat r}=\epsilon_r \sqrt{1-N^2}$, where $\epsilon_r=\pm 1$ for either outgoing ($+$) or ingoing ($-$) radial motion.
These properties fully characterize the the Painlev\'e-Gullstrand observers \cite{Painleve21,Gullstrand22} in the Kerr family of solutions.
The associated refraction index (\ref{nurad}) thus becomes
\beq
\label{nurad2}
n_{{\mathcal N}}=\frac{1}{ 1- \epsilon_r \frac{\sqrt{g_{rr}}}{N}\sqrt{1-N^2} \zeta_{\rm (r)} }
=\frac{1}{ 1-  \sqrt{1-N^2} |\hat \nu^{\hat r}| }\,,
\eeq
implying that Painlev\'e-Gullstrand observers are such that $n_{{\mathcal N}}=1$ either in the case of circularly rotating photons (i.e., $\hat \nu^{\hat r}=0$) or if the spacetime coordinates are such that the lapse function has the fixed value $N=1$.
This is accomplished by suitably transforming the spacetime coordinates so that the new time coordinate is adapted to Painlev\'e-Gullstrand observers. 
For instance, for ingoing radial motion in the Kerr spacetime, the Kerr metric (\ref{kerrfunct1})--(\ref{kerrfunct2}) written in Painlev\'e-Gullstrand adapted coordinates takes the form (see, e.g., Ref. \cite{defbibook} and references therein)
\begin{eqnarray}
\label{PGkerr}
d s^2=&-&\left(1-\frac{2\mathcal{M}r}{\Sigma}\right)d T^2 + 2 \sqrt{\frac{2\mathcal{M}r }{r^2+a^2}} d T d r
\nonumber \\
{}
&-&\frac{4a\mathcal{M}r }{\Sigma}\sin^2\theta\, d T \,d \Phi +\sin^2\theta \left[(r^2+a^2)+\frac{2a^2\mathcal{M}r}{\Sigma}\sin^2\theta \right] d \Phi^2 \nonumber \\
{}
& &
-2a\sin^2\theta\sqrt{\frac{2\mathcal{M}r }{r^2+a^2}} d r d \Phi+\frac{\Sigma}{(r^2+a^2)}d r^2 
+\Sigma d \theta^2\,,
\end{eqnarray}
which has unit lapse (i.e., $1/\sqrt{-g^{TT}}=1$), as it can be easily checked.
Furthermore, it shares the useful property to be regular on the horizon $r=r_+$ (Painlev\'e-Gullstrand coordinates are horizon-penetrating).
The relation with Boyer-Lindquist coordinates $(t,r,\theta,\phi)$ is given by
\beq
\label{PGcoordskerr}
T= t-\int^r f(r) d r\, ,  \qquad \Phi = \phi -\int^r \frac{a}{r^2+a^2} f(r) d r\,,
\eeq
with $r$ and $\theta$ unchanged and  
\beq
f(r)=-\frac{\sqrt{2{\mathcal M}r(r^2+a^2)}}{\Delta}\,.
\eeq
In these new coordinates, Painlev\'e-Gullstrand observers are $T$-slicing-adapted observers, i.e., their four velocity is given by
\beq
{\mathcal N}^\flat= -dT\,.
\eeq
Starting again from the definitions (\ref{dell_def2})--(\ref{coord_time}), the associated refraction index is thus given by
\beq
n_{{\mathcal N}}=1\,,
\eeq
irrespective of the photon path.

Such an enlightening example shows that one cannot simply get the new refraction index as if it were a scalar function by transforming it from the old coordinates to the new ones.
Instead, one must either start again from the relations (\ref{dell_def2})--(\ref{coord_time}) and use the new set of coordinates to define
the new refraction index or consider the metric quantities entering the definition of the refraction index and properly transform them.
Therefore, the construction presented in Section 2 is general enough.

\section{Discussion and concluding remarks}

We have studied the observer-dependent definition of refraction index and optical path for photons moving in a given gravitational background and general  families of observers, generalizing a pioneering approach due to M\"oller \cite{moller}.
Special attention has been devoted to certain families of observers which play a central role in stationary axisymmetric spacetimes (like Schwarzschild and Kerr spacetimes), i.e., static observers, locally nonrotating observers and observers moving along circular orbits.
Interestingly, we have shown that it is always possible to identify a family of observers with respect to whom the refraction index appears equal to unity all along a specific photon path.
 
It is worth to note that this approach does not aim at defining an analogue of the gravitational field in terms of a dielectric (in general anisotropic) medium: this latter interesting topic has received enough attention in the literature by Plebanski \cite{pleb} and Landau and Lifshitz \cite{LL}, and it is mostly used to solve or study Maxwell's equation for a test electromagnetic field on a curved background.

In the approach considered in the present paper only null orbits enter and the arbitrariness in the choice of a parametrization along them is used to set up a \lq\lq formal" analogy with optical path and refraction index.
It is quite interesting that the mentioned analogy may be used to select special observers (i.e., special timelike orbits) starting from the way in which they see null orbits.

The most intriguing effect implied by the equivalence with an optical medium is the possibility for the associated refraction index to become less or equal to 1 either for special values of the spacetime parameters or for special family of observers.
In this situation, the (coordinate) speed of a photon may become greater than the speed of light in vacuum. 
A mathematical explanation is based on the relation $dt\le d\ell_{(P,u)}$, stating that the rate of increasing of the relative length parameter used in the definition of the refraction index is greater than that associated with the coordinate time. Nothing bad in this property, except that it may be the source of counter-intuitive phenomena.

Riding on this equivalent optical medium analogy, however, we found it interesting that
the existence as well as the practical realization of material exhibiting very low (less than 1), even negative and, in general, time-dependent refraction indices is a rather well established fact.
This is the case of the so-called \lq\lq metamaterials" \cite{liu}, i.e., materials 
constructed by using an artificial type of matter, usually a combination of individual materials. 
Their development allows for wide ranging control of material parameters, to the extent of even achieving a negative refraction index.
As a result, the observer-dependent way to look at the spacetime geometry as an equivalent optical medium says that the gravitational field may act not simply as an ordinary material, but also as a metamaterial.

\end{document}